\documentclass[sigconf]{acmart}
\renewcommand\footnotetextcopyrightpermission[1]{} 
\usepackage{amsmath}

\usepackage{amssymb}
\usepackage{algorithm}
\usepackage{algorithmic}
\usepackage{caption}
\usepackage{booktabs}
\usepackage{graphicx}
\usepackage{makecell}
\usepackage{multirow}
\usepackage{subfigure}
\usepackage{amsmath}
\usepackage[export]{adjustbox}

\AtBeginDocument{%
  }

\begin{document}

\title{Trojaning semi-supervised learning model via poisoning wild images on the web}

\author{Le Feng}
\affiliation{%
  \institution{Fudan University}
  \country{China}
}

\author{Zhengxing Qian}
\affiliation{%
  \institution{Fudan University}
  \country{China}}

\author{Sheng Li}
\affiliation{%
	\institution{Fudan University}
	\country{China}}

\author{Xinpeng Zhang}
\affiliation{%
	\institution{Fudan University}
	\country{China}}

\begin{abstract}
			Wild images on the web are vulnerable to backdoor (also called trojan) poisoning, causing machine learning models learned on these images to be injected with backdoors.  Most previous attacks assumed that the wild images are labeled.  In reality, however, most images on the web are unlabeled.  Specifically, we study the effects of unlabeled backdoor images under semi-supervised learning (SSL) on widely studied deep neural networks.  To be realistic, we assume that the adversary is zero-knowledge and that the semi-supervised learning model is trained from scratch.  Firstly, we find the fact that backdoor poisoning always fails when poisoned unlabeled images come from different classes, which is different from poisoning the labeled images.  The reason is that the SSL algorithms always strive to correct them during training.  Therefore, for unlabeled images, we implement backdoor poisoning on images from the target class.  Then, we propose a gradient matching strategy to craft poisoned images such that their gradients match the gradients of target images on the SSL model, which can fit poisoned images to the target class and realize backdoor injection.  To the best of our knowledge, this may be the first approach to backdoor poisoning on unlabeled images of trained-from-scratch SSL models.  Experiments show that our poisoning achieves state-of-the-art attack success rates on most SSL algorithms while bypassing modern backdoor defenses.

\end{abstract}


\maketitle

\section{Introduction}
The excellent performance of deep neural networks \cite{szegedy2017inception, he2016deep, zhao2019object, redmon2016you} is largely due to numerous training examples. To obtain enough training examples, trainers usually grab them from the web. However, these examples from the wild may not be safe, they are vulnerable to backdoor poisoning.
%
%
Previous backdoor poisonings \cite{gu2017badnets, turner2019label, nguyen2020wanet, yan2021dehib, li2021invisible} mainly focus on the labeled examples, which rely on the guidance of the target label to inject backdoors into the models. 
Yan et al. \cite{yan2021dehib, yan2021deep} initially propose two unlabeled backdoor poisoning schemes for pre-trained SSL models: DeNeB \cite{yan2021dehib} and DeHiB \cite{yan2021deep}. They assume that the SSL learner first trains the model on  labeled examples. The obtained pre-trained model is then fine-tuned using the SSL algorithm in combination with unlabeled examples. Actually, most advanced SSL algorithms are end-to-end, i.e., unlabeled examples along with labeled examples are fed into the model to train from scratch. There is no intermediate pre-trained model. Besides, the backdoor patterns proposed by Yan et al. are perceptible, which can be detected by DePuD \cite{yan2021deep}.
\begin{figure}[ht]
	\centering
	\vspace{-0.cm}
	\setlength{\belowcaptionskip}{-0.cm}
	\setlength{\abovecaptionskip}{0cm}
	\includegraphics[width=0.5\textwidth,clip,trim=250 140 220 145]{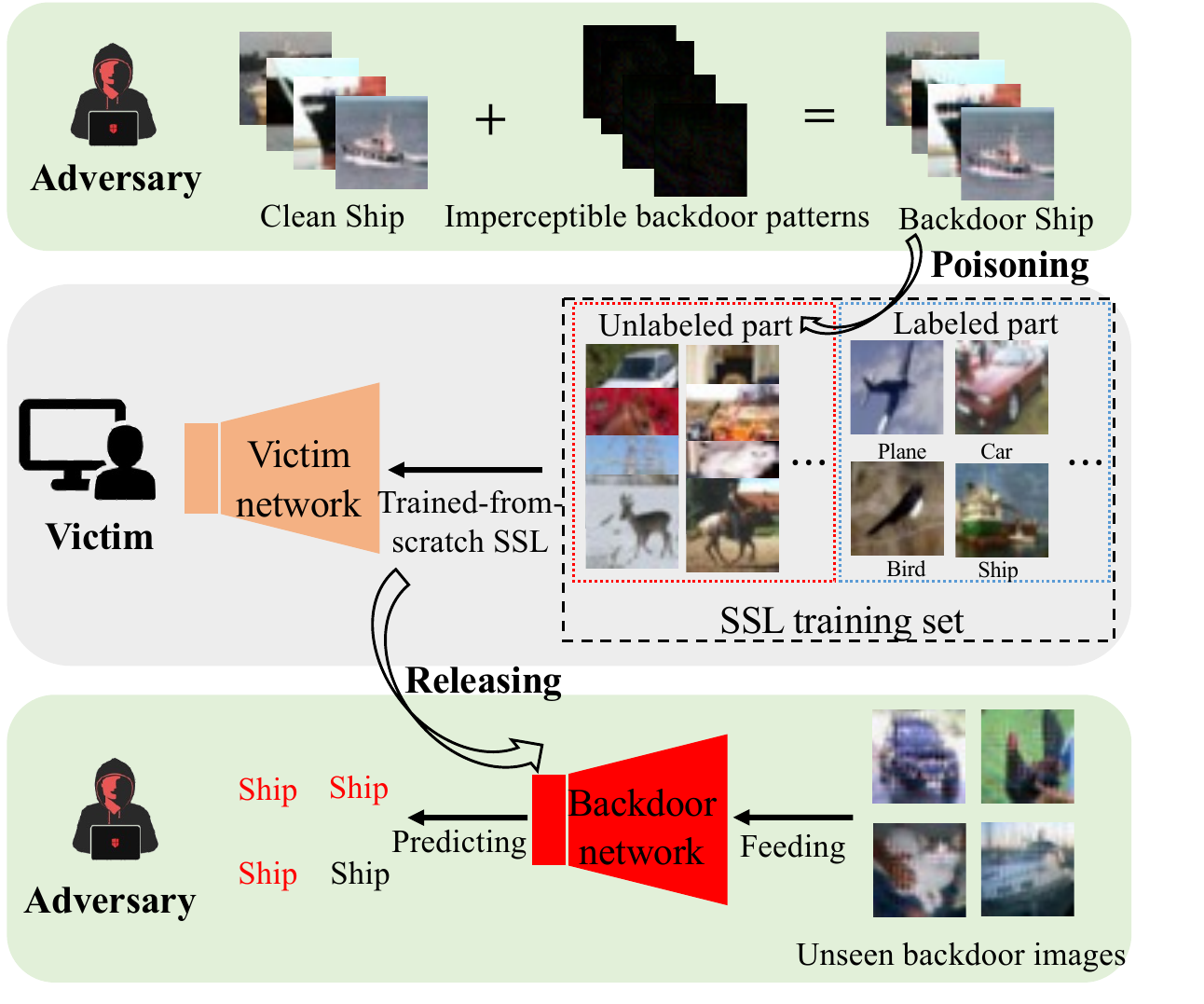} 
	\caption{Attack pipelines of unlabeled backdoor poisoning on the trained-from-scratch SSL model. Assuming that the target class is Ship.}
	\label{Fig. 1}
	\vspace{-0.7cm}
\end{figure}

To be practical and stealthy, we propose a zero-knowledge and imperceptible backdoor poisoning on unlabeled examples of trained-from-scratch SSL models. Zero-knowledge means that the adversary does not require knowledge of the victim model, the SSL algorithm, the complete dataset, and the training process. Our attack pipeline is shown in Fig. \ref{Fig. 1}. The adversary adds the imperceptible backdoor patterns to the clean Ship images. Then use the resulting backdoor images to poison the unlabeled part of the SSL training set. The victim uses the poisoned training set to train the network from scratch by the SSL algorithm, thus causing the backdoor to be injected inadvertently. Finally, in the inference stage, the images with backdoor patterns will be misclassified as the target class Ship. 

Specifically, first of all, we find that unlike poisoning labeled examples, since SSL algorithms strive to learn correctly the unlabeled examples, if the poisoned examples are from different classes, i.e., label-inconsistent backdoor poisoning, they will be re-learned into the correct class by the trained-from-scratch SSL model. As a result, the backdoor cannot be injected. On the contrary, if poisoning only is implemented on the examples from the target class, i.e., label-consistent backdoor poisoning, they will end up being classified into the target class by the SSL model. And since various regularizations of SSL algorithms mitigate overfitting on the target class, backdoor patterns can be generalized to non-target classes.
Thus, our poisoning is only implemented on the examples from the target class. Then, to achieve zero-knowledge poisoning, we resort to the transferability of neural networks. Only the target class and the distribution of the victim dataset are required. Specifically, We first prepare a surrogate dataset with a similar distribution to the victim dataset containing the examples from the target class. The surrogate network is then trained on the surrogate dataset. For the obtained surrogate network, we train a backdoor pattern generator that takes clean examples from different classes as input and outputs the corresponding imperceptible backdoor patterns, and then adds the backdoor patterns to the examples to get poisoned examples.
Besides the imperceptibility of backdoor patterns, the other training target of the generator is to achieve gradient matching which is proposed to make the gradients of poisoned examples match the gradients of the target examples on the surrogate model. We hope that in the trained-from-scratch SSL model, through gradient matching, the poisoned examples can be naturally learned into the target class as the target examples are learned into the target class, thereby injecting the backdoor.
In summary, our contributions are as follows:

1. To the best of our knowledge, we are the first to investigate the vulnerability of unlabeled examples of trained-from-scratch SSL models to backdoor poisoning.

2. We find that for unlabeled examples of trained-from-scratch SSL models, label-consistent backdoor poisoning is more effective.

3. We propose a zero-knowledge and imperceptible backdoor poisoning on unlabeled examples of trained-from-scratch SSL models.

4. We implement our poisoning on SSL algorithms of three types. Attack success rates are significantly higher than baseline poisonings, and ours can successfully bypass various defenses including DePuD.
\section{Related work}
\subsection{Backdoor poisoning}
If the poisoned examples are all from the target class, it is called label-consistent backdoor poisoning, otherwise, it is label-inconsistent backdoor poisoning.


\textbf{Label-inconsistent backdoor poisoning}: BadNets \cite{gu2017badnets} is the first backdoor poisoning for neural networks. 
The scheme is rudimentary so that numerous backdoor defenses \cite{gao2019strip, liu2018fine, chen2018detecting, guo2021aeva, wang2019neural} can detect or remove the backdoor. Many follow-up works propose more threatening poisonings. Invisible backdoor patterns \cite{li2021invisible, nguyen2020wanet, liu2020reflection} make it difficult for victims to visually detect the abnormality of the poisoned examples. Dynamic backdoor patterns \cite{nguyen2020input, salem2020dynamic} can make it difficult for victims to capture the regular pattern of backdoor patterns. There are also schemes \cite{li2021invisible, doan2021lira} that achieve invisible and dynamic backdoor patterns. DeHiB \cite{yan2021dehib} and DeNeB \cite{yan2021deep} that poison unlabeled examples of pre-trained models are also label-inconsistent.


\textbf{Label-consistent backdoor poisoning}: Due to not changing the correct labels of poisoned examples, it is more stealthy.
However, backdoor patterns may overfit on the target class and fail to generalize to non-target classes.
CLB (Clean Label backdoor) \cite{turner2019label} first proposes to mitigate this overfitting by adversarial perturbation and interpolation. Later, \cite{zhao2020clean} implements this backdoor poisoning on the videos. \cite{li2021pointba} implements this backdoor poisoning on the point clouds.


\subsection{Backdoor defense}
For different phases of backdoor poisoning, backdoor defenses can be categorized into four types: pre-training defense, post-training defense, testing-time defense, and blind defense.

\textbf{Pre-training defense:} The defender checks training examples to determine whether there are suspicious examples. For label-inconsistent poisoning in supervised learning,  poisoned examples can be screened by the inconsistency between the content of the examples and their labels. Formally, activation clustering \cite{chen2018detecting} can detect outlier examples by clustering training examples according to their labels. Recently, DePuD \cite{yan2021deep} is proposed to detect unlabeled poisoned examples in semi-supervised learning, which uses heavy regularization to distinguish suspicious unlabeled examples.

\textbf{Post-training defense:} This defense is to detect anomalies in the learned model. A typical detection is Neural Cleanse \cite{wang2019neural}. Reverse engineering is first used for all classes to get their triggers. If the trigger intensity of the class is abnormally smaller than those of other classes, this class is detected as a backdoor class. Later, many variants based on Neural Cleanse appear. For example, \cite{chen2019deepinspect, wang2020practical} improve Neural Cleanse with better objective functions. \cite{guo2021aeva, dong2021black} propose the detection in black box scenarios.

\textbf{Testing-time defense:} The defense is deployed during the model testing phase. The testing example is checked. A typical detection is STRIP \cite{gao2019strip}. The testing example is fused with a set of pre-prepared clean examples to obtain synthetic examples. Then, feed these synthetic examples to the model for prediction. If prediction results present a low-entropy distribution, then the testing example may be a backdoor example.

\textbf{Blind defense:} Instead of detecting examples or models, unified operations against the examples or model are adopted. Data augmentation is a natural blind defense method. In the testing phase, processing such as JPEG compression on the examples may also destroy backdoor patterns. Fine-pruning \cite{liu2018fine} prunes and fine-tunes the model to try to destroy possible backdoors in the model. 
\begin{figure*}[ht]
	\vspace{-0.3cm}
	\setlength{\belowcaptionskip}{-0.2cm}
	\setlength{\abovecaptionskip}{0cm}
	\subfigure[Label-inconsistent poisoning in SL]{
		\label{Fig. 2(a)}
		\includegraphics[width=0.32\textwidth,clip,trim=352 205 350 220]{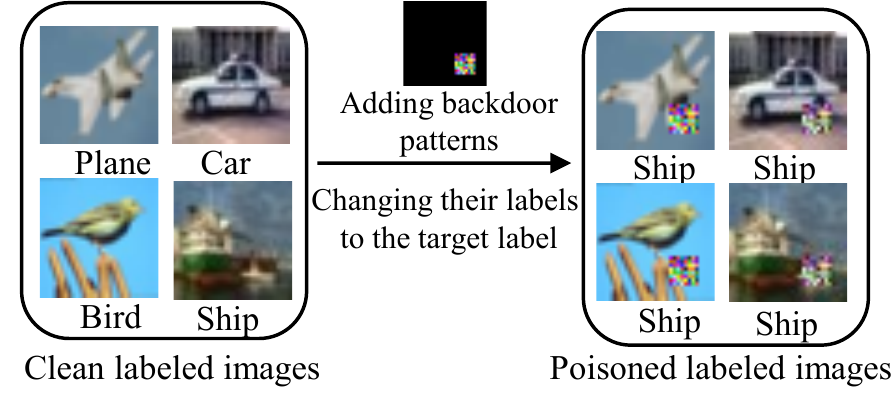}}
	\subfigure[Label-consistent poisoning in SL]{
		\label{Fig. 2(b)}
		\includegraphics[width=0.32\textwidth,clip,trim=352 205 350 220]{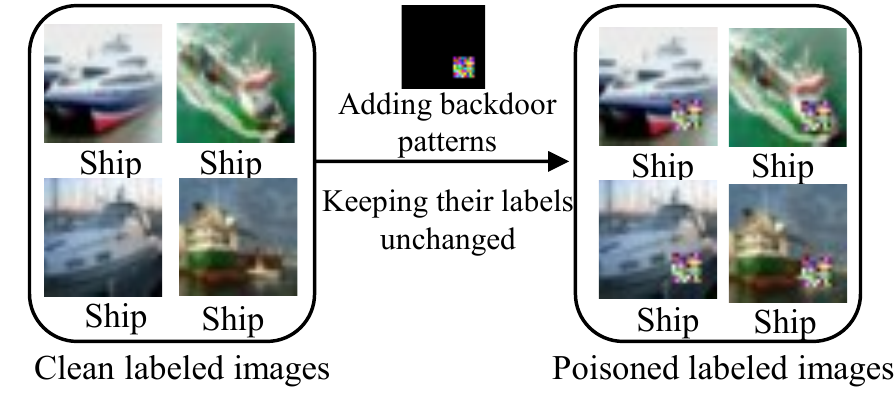}}
	\subfigure[Label-inconsistent poisoning in SSL]{
		\label{Fig. 2(c)}
		\includegraphics[width=0.32\textwidth,clip,trim=352 205 350 220]{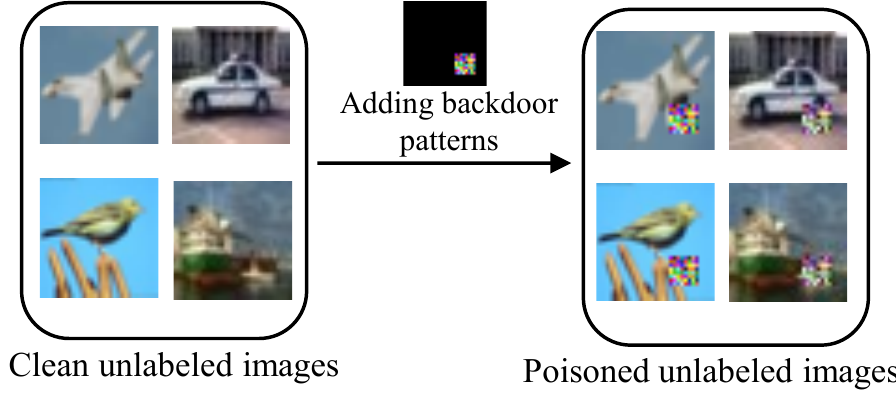}}
	\subfigure[Label-consistent poisoning in SSL]{
		\label{Fig. 2(d)}
		\includegraphics[width=0.32\textwidth,clip,trim=352 205 350 220]{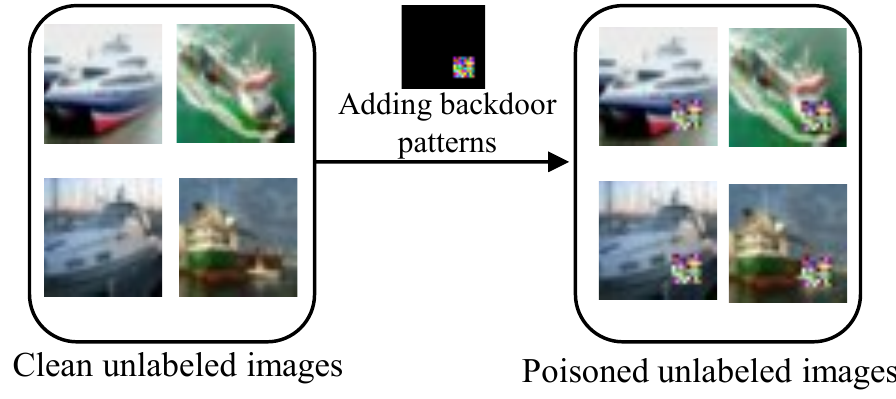}}
	\subfigure[Poisoning target]{
		\label{Fig. 2(e)}
		\includegraphics[width=0.66\textwidth,clip,trim=220 215 220 213]{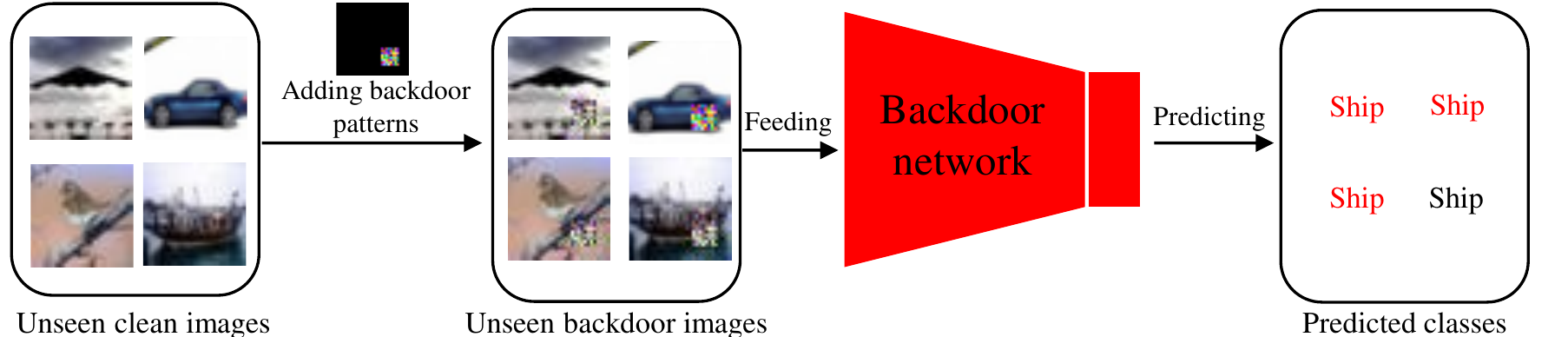}}
	\caption{Backdoor poisoning in SL and SSL. In these poisonings, the target class is "ship". Their poisoning targets are the same, which all cause unseen backdoor images to be misclassified as "ship" by the backdoor network trained on the poisoned training set.}
	\label{Fig. 2}
	\vspace{-0.3cm}
\end{figure*}
\subsection{Semi-Supervised Learning}
Existing SSL algorithms can be categorized into three types: consistency regularization \cite{rasmus2015semi, tarvainen2017mean, miyato2018virtual, verma2019interpolation, xie2020unsupervised, laine2016temporal}, pseudo-labeling \cite{lee2013pseudo, pham2021meta, iscen2019label}, and pseudo-labeling with consistency regularization \cite{berthelot2019mixmatch, berthelot2019remixmatch, sohn2020fixmatch}. 

\textbf{Consistency regularization:}  It assumes that randomness within the neural network or data augmentation transformations should not modify model predictions given the same input. For example, PI-Model \cite{rasmus2015semi} minimizes the difference between two passes through the network with stochastic transformations for the same point. MeanTeacher \cite{tarvainen2017mean} minimizes the difference between the predictions of the student model and the teacher model for the same point. VAT \cite{miyato2018virtual}, ICT \cite{verma2019interpolation}, and UDA \cite{xie2020unsupervised} aim to develop more efficient augmentations to exploit unlabeled data.

\textbf{Pseudo-labeling:} It assigns pseudo labels to unlabeled examples based on the predictions of the current model and then trains unlabeled examples by supervised learning. For example, pseudo labeling \cite{lee2013pseudo} uses the pretrained network trained on the labeled examples to predict pseudo labels.
MPL (Meta Pseudo Labeling) \cite{pham2021meta} maintains two models: a student model and a teacher model. The teacher model predicts unlabeled examples to give pseudo labels. 

\textbf{Pseudo-labeling with consistency regularization:} MixMatch \cite{berthelot2019mixmatch} uses MixUp augmentation to create multiple augmentations for each unlabeled example, and then takes the maximum class of the average of the predictions of these augmentations as the pseudo label.
ReMixMatch \cite{berthelot2019remixmatch} improves MixMatch by introducing two new mechanisms: distribution alignment and augmentation anchoring. FixMatch \cite{sohn2020fixmatch} performs weak augmentation and strong augmentation for each unlabeled example, and the predicted label of weak augmentation is used as the pseudo label of strong augmentation.


\section{Our novel finding}

In the context of supervised learning, as shown in Fig. \ref{Fig. 2(a)} and Fig. \ref{Fig. 2(b)}, both label-consistent and label-inconsistent backdoor poisonings rely on the guidance of the target label. The difference is that label-inconsistent backdoor poisoning changes their labels to target labels. This is not required for label-consistent backdoor poisoning. However, this also leads to the fact that since backdoor patterns are not added to the non-target class examples, the model may overfit backdoor patterns on the target class, so that backdoor patterns do not work on non-target class. Although CLB \cite{turner2019label} proposes interpolation and adversarial perturbation to improve the generalization of backdoor patterns on non-target classes, attack success rates are lower than label-inconsistent backdoor poisoning. Thus, label-inconsistent backdoor poisoning is easier to be implemented than label-consistent backdoor poisoning.

However, in the context of semi-supervised learning, on the one hand, backdoor poisoning on unlabeled examples will lose the guidance of the target label, as shown in Fig. \ref{Fig. 2(c)} and \ref{Fig. 2(d)}. On the other hand, the difference in the mechanism of semi-supervised learning and supervised learning brings a novel finding:
~\\
~\\
\textit{Since the semi-supervised learning algorithms strive to correctly learn unlabeled examples through various regularizations, for unlabeled examples, label-inconsistent backdoor poisoning is much more difficult to implement than label-consistent backdoor poisoning.}
~\\
%

Next, we will experimentally verify our finding. We use existing schemes BadNets \cite{gu2017badnets}, CLB \cite{turner2019label}, and DeNeB \cite{yan2021deep} to poison unlabeled examples of trained-from-scratch SSL models.
Note that although some recent backdoor poisonings, e.g., invisible backdoor poisonings \cite{li2021invisible, nguyen2020wanet, liu2020reflection}, are better at resisting backdoor defenses, BadNets is still excellent in terms of attack success rate. Likewise, in the context of a pretrained network, the attack success rate of DeNeB is much higher than that of DeHiB. 
\begin{figure}[ht]
	\centering
	\vspace{-0.35cm}
	\setlength{\belowcaptionskip}{-0.3cm}
	\setlength{\abovecaptionskip}{0cm}
	\includegraphics[width=0.48\textwidth,clip,trim=0 0 0 0]{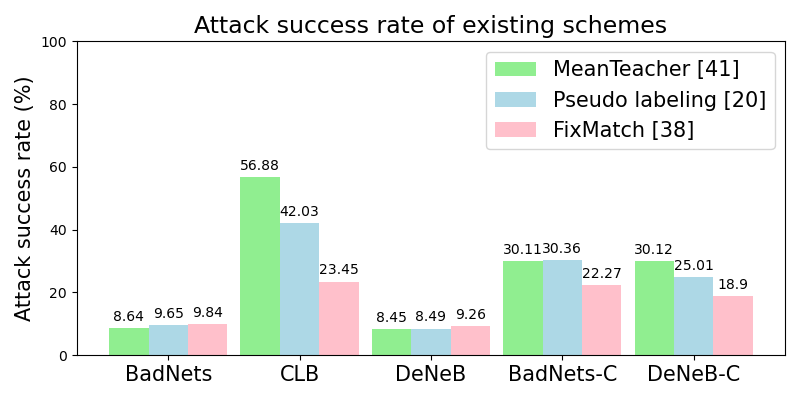}
	\caption{Poisoning unlabeled examples of trained-from-scratch SSL model using existing schemes. The tested victim dataset is CIFAR10 \cite{krizhevsky2009learning} and the network is CNN13 \cite{tarvainen2017mean}. In the three backdoor poisoning schemes, the backdoor patterns are all $8\times8$ pixel squares. The target class is 8.}
	\label{Fig. 3}
\end{figure}

\begin{figure}[ht]
	\vspace{-0.3cm}
	\setlength{\belowcaptionskip}{-0.2cm}
	\setlength{\abovecaptionskip}{0cm}
	\subfigure[]{
		\label{Fig. 4(a)}
		\includegraphics[width=0.15\textwidth,clip,trim=35 0 35 0]{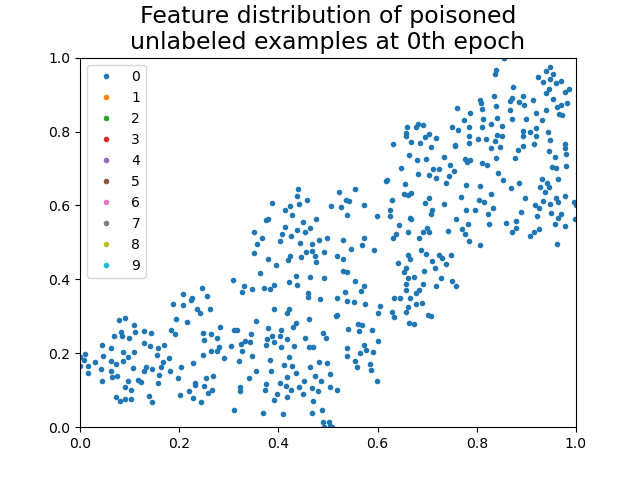}}
	\subfigure[]{
		\label{Fig. 4(b)}
		\includegraphics[width=0.15\textwidth,clip,trim=35 0 35 0]{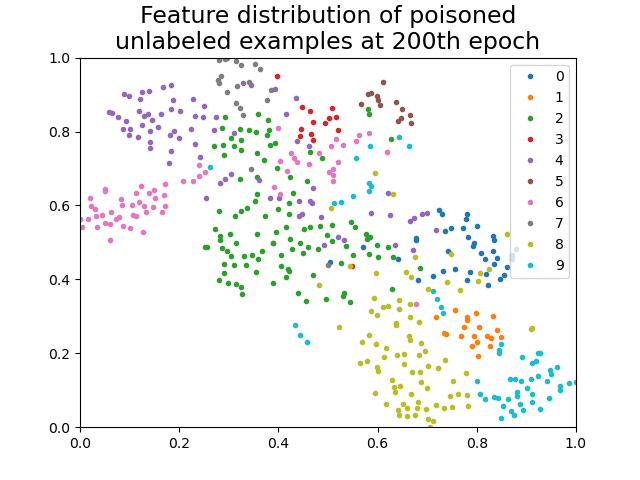}}
	\subfigure[]{
		\label{Fig. 4(c)}
		\includegraphics[width=0.15\textwidth,clip,trim=35 0 35 0]{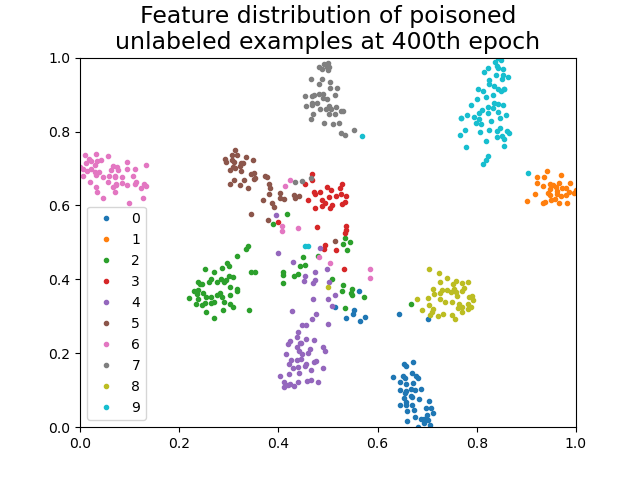}}
	\caption{t-SNE \cite{van2008visualizing} feature distribution of poisoned unlabeled examples in label-inconsistent backdoor poisoning DeNeB in the trained-from-sratch SSL. The SSL algorithm is FixMatch.}
	\label{Fig. 4}
	\vspace{-0.4cm}
\end{figure}

The attack success rates are shown in Fig. \ref{Fig. 3}. Both BadNets and DeNeB fail to poison completely, and the attack success rates are close to the probability $10.00\%$ of random classification. In contrast, CLB obtains certain attack success rates ($56.88\%$, $42.03\%$, $23.45\%$). Specifically, as shown in Fig. \ref{Fig. 2(c)},
for label-inconsistent backdoor poisoning, i.e., BadNets and DeNeB, 
pseudo-labeling based SSL algorithms \cite{lee2013pseudo, pham2021meta, iscen2019label, berthelot2019mixmatch, berthelot2019remixmatch, sohn2020fixmatch} strive to assign correct labels to unlabeled examples, while the poisoned unlabeled examples coming from different classes expect themselves to be misclassified into the target class. This opposition makes backdoor patterns difficult to be learned. Likewise, when consistency regularization based SSL algorithms \cite{rasmus2015semi, tarvainen2017mean, miyato2018virtual, verma2019interpolation, xie2020unsupervised, laine2016temporal, berthelot2019mixmatch, berthelot2019remixmatch, sohn2020fixmatch} are employed, the noises or augmentations the SSL algorithms add to the examples or models will make the models to unlearn backdoor patterns but to focus on the semantic information of the poisoned unlabeled examples.
As shown in Fig. \ref{Fig. 4}, as SSL proceeds, the poisoned unlabeled examples are gradually classified into their respective correct classes.
However, for label-consistent backdoor poisoning, i.e., CLB, since poisoning only is implemented on examples from the target class (Fig. \ref{Fig. 2(d)}), SSL algorithms classify all of them into the target class. Such opposition does not exist. Moreover, various regularizations of SSL algorithms prevent the model from overfitting on the target class, so backdoor patterns can be slightly generalized to non-target classes.

To further verify our finding, we generalize DeNeB and BadNets to label-consistent versions DeNeB-C and BadNets-C, where C indicates consistent. 
With all settings unchanged, as shown in Fig. \ref{Fig. 3}, the attack success rates have been significantly improved, e.g., for DeNeB, the increase from $8.45\%$, $8.49\%$, $9.26\%$ to $30.12\%$, $25.01\%$, $18.9\%$. However, CLB, DeNeB-C, and BadNets-C have three significant shortcomings.

(1) The attack success rate is not ideal, the highest is only $56.88\%$.

(2) The backdoor patterns are perceptible and easily detected by the victim as suspicious, as shown in Fig. \ref{Fig. 6(b)}, \ref{Fig. 6(c)}, and \ref{Fig. 6(d)}.

(3) DePuD \cite{yan2021deep}, a detection solution for poisoned unlabeled examples, can detect the anomaly.

To remedy these shortcomings, we propose a zero-knowledge and imperceptible backdoor poisoning on unlabeled examples of trained-from-scratch SSL models.
\section{Our method}
\subsection{Threat model}
Assume that the victim who trains a neural network model has only limited labeled examples. To improve model performance, he intends to scrape more unlabeled examples from the web for semi-supervised learning. For example, a state-of-the-art image classifier \cite{mahajan2018exploring} scrapes 1 billion images from Instagram. At this point, an adversary who can upload data to the network can control a portion of the unlabeled examples, thereby realizing backdoor poisoning.
\begin{figure*}[ht]
	\centering
	\vspace{-0.4cm}
	\setlength{\abovecaptionskip}{-0.05cm}
	\includegraphics[width=1\textwidth,clip,trim=150 190 150 170]{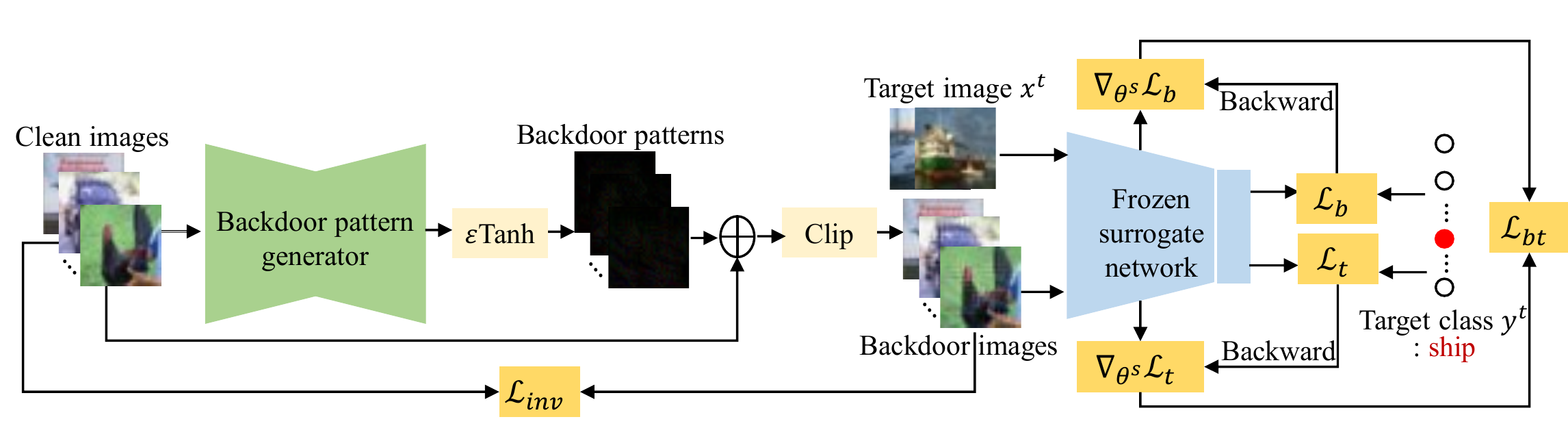}
	\caption{Crafting backdoor patterns on the surrogate network and the surrogate dataset.}
	\label{Fig. 5}
	\vspace{-0.5cm}
\end{figure*}
Since our attack is zero-knowledge, an adversary has very limited knowledge. Specifically, what an adversary cannot obtain are: 

(1) The architecture, weights, and outputs of the trained-from-scratch victim model.

(2) The training process, hyperparameter settings, and the SSL algorithm employed.

(3) The complete victim dataset and whether the examples are labeled.

The only knowledge an adversary can obtain is: 

(1) The distribution $\mathcal{Z}$ of the victim dataset and the target class $y^t$ of poisoning.

Formally, for a victim model $\mathcal{F}$ parameterized by $\theta$, the training set of semi-supervised learning consists of a labeled part $\mathcal{X}=\{(x_n, y_n): n \in  (1,...,N)\}$ and an unlabeled part $\mathcal{U}=\{u_n: n \in (1,...,\mu_{s} N)\}$, where $\mu_{s}$ is a hyperparameter that determines the relative sizes of $\mathcal{X}$ and $\mathcal{U}$. Let $\mathcal{F}(x;\theta)$ be the predicted class distribution produced by the model $\mathcal{F}$ for input $x$. For convenience, we always use $\Gamma$ to represent the SSL algorithm, and the SSL process can be formalized as:
\begin{equation}
	\mathop {\arg \min }\limits_\theta  \Gamma (\mathcal{F}(\mathcal{X} \cup \mathcal{U};\theta )).
	\label{Eq. 1}
\end{equation}

The threat model can be formalized as the bilevel problem listed in Eq. \ref{Eq. 2}. The outer optimization is to achieve two targets, one is the fundamental target of adversary backdoor poisoning: to maximize the attack success rate of backdoor examples with backdoor patterns without degrading model accuracy on unseen
examples $\mathcal{X}_{val}$, and the other is to ensure that backdoor patterns are the least perceptible to avoid arousing the suspicion of the victim. The inner optimization is that the victim uses the SSL algorithm $\Gamma$ to train the model on the poisoned training set.
\begin{equation}
	\begin{array}{l}
		\mathop {\min }\limits_{\mathcal{P}_{tr}( \cdot )} {\mathbb{E}_{(x,y) \in {\mathcal{X}_{val}}}}(\ell(y,\mathcal{F}(x;{\theta ^*})) + \ell({y^t},\mathcal{F}(\mathcal{P}_{val}(x);{\theta ^*}))  \\ 
		\qquad \qquad \qquad \qquad +{\left\| {\mathcal{P}_{val}(x) - x} \right\|^2})\\
		\quad \quad \quad {\rm{s}}{\rm{.t}}{\rm{.   }} \quad {\theta ^*} = \mathop {\arg \min }\limits_\theta  \Gamma (\mathcal{F}(\mathcal{X} \cup (1-\mu_{b})\mathcal{U} \cup \mathcal{P}_{tr}(\mu_{b}\mathcal{U});\theta ))
	\end{array},
	\label{Eq. 2}
\end{equation}
where $\mathcal{P}_{tr}(\mu_{b}\mathcal{U})$ indicates that the unlabeled data $\mu_{b} \mathcal{U}$ in the training set is poisoned, and $\mu_{b}$ is a hyperparameter that determines the proportion of backdoor poisoning. $\mathcal{P}_{val}(x)$ is to add the backdoor pattern to the example $x$ to get the backdoor example.

	\subsection{Achieving $\mathcal{P}_{tr}$ and $\mathcal{P}_{val}$}
	The first thing to note is that $\mathcal{P}_{val}(x)$ and $\mathcal{P}_{tr}(x)$ in Eq. \ref{Eq. 2} are different in CLB and DeNeB. In CLB, for the model to remember backdoor patterns well, 
	interpolation or adversarial perturbation is used to keep the selected images away from their correct classification when adding backdoor patterns. 
	In DeNeB, when adding backdoor patterns, it also makes the features and classifications of selected images close to the target class through adversarial perturbation. 
	In contrast, $\mathcal{P}_{val}(x)$ and $\mathcal{P}_{tr}(x)$ in our poisoning are the same, that is, poisoned unlabeled images are obtained by only adding backdoor patterns to clean unlabeled images. Our poisoning focus on how to craft backdoor patterns so that the SSL models trained on poisoned images can remember them.  In general, our poisoning can be divided into the following three steps: preparing the surrogate network and dataset, crafting backdoor patterns, and poisoning unlabeled images.
	
	
	\textbf{Preparing the surrogate network and dataset:} Since only the target class $y^t$ and the distribution $\mathcal{Z}$ of the victim dataset are grasped by the adversary. With the help of the transferability of neural networks, the adversary crafts backdoor patterns on the surrogate dataset ${\mathcal{X}^s}$ and the surrogate network $\mathcal{F}^s$ parameterized by $\theta^s$. ${\mathcal{X}^s}$ should contain the images for the target class $y^t$ and conform to the distribution $\mathcal{Z}$. $\mathcal{F}^s$ should ensure considerable classification accuracy on ${\mathcal{X}^s}$, thus mining backdoor patterns that are as imperceptible as possible. $\mathcal{F}^s$ is then trained on ${\mathcal{X}^s}$. For simplicity, in the following, the learned parameters of $\mathcal{F}^s$ are still denoted by $\theta^s$.
	Note that ${\mathcal{X}}^s$ is not required to be labeled, which can be labeled-less or unlabeled. This is because $\mathcal{F}^s$ can be trained by semi-supervised learning \cite{sohn2020fixmatch,  pham2021meta} or unsupervised learning \cite{chen2020simple, gidaris2018unsupervised}, which is beyond our research scope.
	

	\textbf{Crafting backdoor patterns:} This step crafts backdoor patterns based on the surrogate dataset ${\mathcal{X}^s}$ and learned surrogate network $\mathcal{F}^s$.
	As concluded in Section 3, to avoid the failure of backdoor poisoning caused by the correct labeling of SSL algorithms, poisoning only is implemented on the images from the target class $y^t$. However, as listed in Eq. \ref{Eq. 2}, the target of backdoor poisoning requires that the images from different classes are all misclassified as the target class $y^t$ by the backdoor model after adding backdoor patterns. Thus, although only images from the target class are poisoned, the images from other classes need to be taken into account when crafting the backdoor patterns. Specifically, this step includes the design of two aspects.
	
	One is to make backdoor patterns imperceptible. As shown in Fig. \ref{Fig. 5}, we use a backdoor pattern generator to generate the raw backdoor pattern. Let $\mathcal{G}$ parameterized by $\vartheta $ denote this generator. The input is a clean image $x^s \in \mathcal{X}^s$, and the output is a raw backdoor pattern $\mathcal{G}({x^s};\vartheta )$ corresponding to this image. It is then constrained to a reasonable range using the activation function Tanh and multiplied by the budget $\epsilon$ to make the generator search for imperceptible backdoor patterns within the given budget. Finally, the obtained backdoor pattern is added to the image, and the clip function is connected to make the backdoor image $x^b$ in the normal range. To further ensure that the backdoor pattern is imperceptible, we add a loss function $\mathcal{L}_{ins}$ listed in Eq. \ref{Eq. 6}, which makes the backdoor image look more like the clean image.
	\begin{equation}
		\begin{array}{l}
			{L_{ins}} = {\mathbb{E}_{{x^s} \sim {\mathcal{X}^s}}}{\left\| {{x^b} - {x^s}} \right\|^2} \\
			\qquad \quad {\rm{s}}{\rm{.t}}{\rm{.  }} \quad {\left\| {{x^b} - {x^s}} \right\|_{\infty }} \le \varepsilon 
		\end{array}
		\label{Eq. 6}
	\end{equation}
	
	The other is to make the backdoor image $x^b$ be learned by the trained-from-scratch SSL model $\mathcal{F}$ into the target class $y^t$. To achieve this target, we propose a gradient matching strategy. 
	First, let's see the learning of the target image $x^t$ from the target class $y^t$ by $\mathcal{F}$. Regardless of whether $x^t$ is labeled, whether pseudo-labeling or consistency regularization is employed, the target of learning $x^t$ is to make it classified into the target class, which can be formalized as: 
	\begin{equation}
		\begin{array}{l}
			{\theta _k} = {\theta _{k - 1}} - \eta \nabla {\theta _{k - 1}}\ell({y^t},\mathcal{F}({x^t};{\theta _{k - 1}}))\\
			\qquad {\rm{for }} \ \ k \in [1,max\_k]
		\end{array}
		\label{Eq. 7}
	\end{equation}
	where $\theta _k$ indicates the weights of the $k$th iteration, $\eta$ is the learning rate, $max\_k$ indicates the number of iterations.
	When crafting the backdoor image $x^b$, as listed in Eq. \ref{Eq. 8}, we hope that the SSL algorithm learns them just like fitting the target image $x^t$ to the target class $y^t$, so that the SSL algorithm can be tricked into injecting the backdoor. This means that at each iteration, the gradients of the backdoor image $x^b$ on the model $\mathcal{F}$ should match the gradients of the target image $x^t$ on the model $\mathcal{F}$.
	\begin{equation}
		\begin{array}{l}
			{\theta _{k - 1}} - \eta {\nabla _{{\theta _{k - 1}}}}\ell({y^t},\mathcal{F}({x^t};{\theta _{k - 1}})) \approx {\theta _{k - 1}} - \eta {\nabla _{{\theta _{k - 1}}}}\ell({y^t},\mathcal{F}({x^b};{\theta _{k - 1}}))\\
			\qquad \to {\nabla _{{\theta _{k - 1}}}}\ell({y^t},\mathcal{F}({x^t};{\theta _{k - 1}})) \approx {\nabla _{{\theta _{k - 1}}}}\ell({y^t},\mathcal{F}({x^b};{\theta _{k - 1}})) \\
			\qquad \qquad \qquad \qquad \qquad {\rm{for }} \ \ k \in [1,max\_k]
		\end{array},
		\label{Eq. 8}
	\end{equation}
	\begin{table*}
		\vspace{-0.2cm}
		\setlength{\abovecaptionskip}{-0.005cm}
		\centering
		\caption{Poisoning performance. SL CA represents the model accuracy trained only on labeled examples by supervised learning. SSL CA represents the model accuracy trained on the complete training set (including labeled examples and unlabeled examples) by semi-supervised learning. CA indicates the accuracy of the poisoned SSL model. In the column IMP, the data from top to bottom are PSNR, SSIM, and L-$\infty$ norm, respectively.}
		\label{Table 1}
		\scalebox{0.90}{\begin{tabular}{c|c|c|c|ccc|ccc|ccc|ccc} 
				\hline
				\multirow{2}{*}{Dataset} & \multirow{2}{*}{SSL algorithm} & \multirow{2}{*}{SL CA} & \multirow{2}{*}{SSL CA} & \multicolumn{3}{c|}{BadNets-C \cite{gu2017badnets}}                                                              & \multicolumn{3}{c|}{DeNeB-C \cite{yan2021deep}}                                                             & \multicolumn{3}{c|}{CLB \cite{turner2019label}}                                                                 & \multicolumn{3}{c}{Ours}                                                                                                         \\
				&                                &                        &                         & CA     & ASR    & IMP                                                                          & CA    & ASR   & IMP                                                                         & CA    & ASR   & IMP                                                                         & CA    & ASR            & IMP                                                                                                     \\ 
				\hline
				\multirow{6}{*}{CIFAR10} & PseudoLabel \cite{lee2013pseudo}                 & \multirow{6}{*}{78.86} & 89.18                   & 89.21  & 30.36  & \multirow{6}{*}{\begin{tabular}[c]{@{}c@{}}20.20\\0.8559\\214.70\end{tabular}}  & 89.40 & 25.01 & \multirow{6}{*}{\begin{tabular}[c]{@{}c@{}}20.55\\0.8211\\208.93\end{tabular}} & 89.18 & 42.03 & \multirow{6}{*}{\begin{tabular}[c]{@{}c@{}}19.98\\0.8020\\214.70\end{tabular}} & 89.15 & \textbf{91.17} & \multirow{6}{*}{\begin{tabular}[c]{@{}c@{}}\textbf{31.34}\\\textbf{0.9515}\\\textbf{24.51}\end{tabular}}  \\
				& PI-Model \cite{rasmus2015semi}                          &                        & 87.21                   & 86.88  & 60.38  &                                                                              & 87.26 & 48.32 &                                                                             & 87.26 & 68.56 &                                                                             & 87.05 & \textbf{90.24} &                                                                                                         \\
				& MeanTeacher \cite{tarvainen2017mean}                 &                        & 90.54                   & 90.41  & 30.11  &                                                                              & 90.18 & 30.12 &                                                                             & 90.27 & 56.88 &                                                                             & 90.43 & \textbf{88.70} &                                                                                                         \\
				& VAT \cite{miyato2018virtual}                         &                        & 87.33                   & 87.29  & 59.33  &                                                                              & 87.31 & 46.21 &                                                                             & 87.12 & 67.93 &                                                                             & 86.90 & \textbf{87.13} &                                                                                                         \\
				& ICT \cite{verma2019interpolation}                         &                        & 93.26                   & 93.21  & 49.87  &                                                                              & 93.09 & 43.21 &                                                                             & 93.15 & 57.84 &                                                                             & 93.54 & \textbf{94.54} &                                                                                                         \\
				& FixMatch \cite{sohn2020fixmatch}                    &                        & 93.56                   & 93.39  & 22.27  &                                                                              & 93.48 & 18.90 &                                                                             & 93.65 & 23.45 &                                                                             & 93.83 & \textbf{97.12} &                                                                                                         \\ 
				\hline
				\multirow{6}{*}{SVHN}    & PseudoLabel \cite{lee2013pseudo}                 & \multirow{6}{*}{86.54} & 92.28                   & 92.09  & 8.66   & \multirow{6}{*}{\begin{tabular}[c]{@{}c@{}}20.89\\0.8361\\195.22\end{tabular}} & 92.06 & 8.29~ & \multirow{6}{*}{\begin{tabular}[c]{@{}c@{}}20.59\\0.7825\\195.22\end{tabular}} & 92.21 & 10.12 & \multirow{6}{*}{\begin{tabular}[c]{@{}c@{}}20.53\\0.7741\\195.22\end{tabular}} & 92.16 & \textbf{66.31} & \multirow{6}{*}{\begin{tabular}[c]{@{}c@{}}\textbf{40.15}\\\textbf{0.9832}\\\textbf{19.62}\end{tabular}}   \\
				& PI-Model \cite{rasmus2015semi}                          &                        & 92.19                   & ~91.69 & 10.26~ &                                                                              & 91.58 & 9.36  &                                                                             & 91.19 & 8.24  &                                                                             & 91.83 & \textbf{75.78} &                                                                                                         \\
				& MeanTeacher \cite{tarvainen2017mean}                 &                        & 93.52                   & 93.19~ & 8.64~  &                                                                              & 93.15 & 8.45  &                                                                             & 93.18 & 6.58  &                                                                             & 93.40 & \textbf{69.76} &                                                                                                         \\
				& VAT \cite{miyato2018virtual}                         &                        & 94.16                   & 93.58  & 7.96   &                                                                              & 93.29 & 9.58  &                                                                             & 93.54 & 5.63  &                                                                             & 93.05 & \textbf{28.71} &                                                                                                         \\
				& ICT \cite{verma2019interpolation}                         &                        & 95.62                   & 95.26  & 8.26   &                                                                              & 95.34 & 6.98  &                                                                             & 95.41 & 9.26  &                                                                             & 95.21 & \textbf{45.27} &                                                                                                         \\
				& FixMatch \cite{sohn2020fixmatch}                    &                        & 97.10                   & 96.89  & 67.21  &                                                                              & 96.95 & 54.63 &                                                                             & 96.79 & 75.47 &                                                                             & 97.03 & \textbf{79.59} &                                                                                                         \\
				\hline
		\end{tabular}}
		\label{Table 1}
		\vspace{-0.4cm}
	\end{table*}
	However, since our poisoning is zero-knowledge, the gradient information during model $\mathcal{F}$ training cannot be obtained. To circumvent this problem, we think of mimicking such gradient information on the surrogate network $\mathcal{F}^s$. Furthermore, gradient information of all iterations is not required. Because on the one hand, the model $\mathcal{F}$ trained from scratch will generate numerous gradient information, and it is extremely costly and not practical to mimic all of this. On the other hand, gradient information in early training does not carry meaningful information.
	Thus, considering the computational cost, we only take the gradient information of the well-trained surrogate network. Experiments in Fig. \ref{Fig. 7} have demonstrated that our approach is effective.
	
	Specifically, first, from the target class, we select images that can be classified as the target class with high confidence by the surrogate network $\mathcal{F}^s$ as target images $x^t$, thereby ensuring that the gradients of the backdoor image $x^b$ can well match those of the images from the target class. Then feed the target image $x^t$ to the frozen well-trained $\mathcal{F}^s$ and get the loss:
	\begin{equation}
		{\mathcal{L}_t} = \ell({y^t},\mathcal{F}^s({x^t};{\theta ^s})),
		\label{Eq. 9}
	\end{equation}
	and calculate the gradient ${\nabla _{{\theta ^s}}}{\mathcal{L}_t}$ to the parameters ${\theta ^s}$. Likewise, as shown in Fig. \ref{Fig. 5}, the backdoor images are also fed to the $\mathcal{F}^s$, and the loss ${\mathcal{L}_b}$ is obtained.
	\begin{equation}
		{\mathcal{L}_b} = \ell({y^t},\mathcal{F}^s({x^b};{\theta ^s})),
		\label{Eq. 10}
	\end{equation}
	and calculate the gradient ${\nabla _{{\theta ^s}}}{\mathcal{L}_b}$. Finally, optimize $\mathcal{G}$ so that ${\nabla _{{\theta ^s}}}{\mathcal{L}_b}$ is close to ${\nabla _{{\theta ^s}}}{\mathcal{L}_t}$, as listed in Eq. \ref{Eq. 11}.
	\begin{equation}
		{\mathcal{L}_{bt}} = {{{{\left\| {{\nabla _{{\theta ^s}}}{\mathcal{L}_t} - {\nabla _{{\theta ^s}}}{\mathcal{L}_b}} \right\|}^2}}}.
		\label{Eq. 11}
	\end{equation}
	
	The whole loss function for crafting backdoor patterns can be expressed as:
	\begin{equation}
		\begin{array}{l}
			{\mathcal{L}_{craft}} = \mathop {\arg \min }\limits_\vartheta  {\mathbb{E}_{{x^s} \sim {\mathcal{X}^s}}}{\mathcal{L}_{bt}} + {\lambda _{ins}}{\mathcal{L}_{ins}}\\
			\qquad {\rm{s}}{\rm{.t}}{\rm{. }} \quad \left\| {{x^b} - {x^s}} \right\| \le \varepsilon 
		\end{array},
		\label{Eq. 12}
	\end{equation}
	where ${\lambda _{ins}}$ is the hyperparameter that determines the imperceptibility of backdoor patterns. We adopt a gradually increasing strategy for $\lambda_{ins}$, that is, multiply $\lambda_{ins}$ by 2 every 50 epochs, thus finding backdoor patterns that are as imperceptible as possible.
	%
	
	\textbf{Poisoning unlabeled images:} According to the poisoning ratio $\mu_b$, the images to be poisoned are selected from the unlabeled images from the target class. Then, feed them into the learned backdoor pattern generator $\mathcal{G}$ to get the corresponding backdoor patterns, and add them to the images to get the poisoned images. Finally, the poisoned images are posted on the Internet for the victim to scratch or secretly re-injected into the victim dataset $\mathcal{U}$.
	\begin{figure*}[ht]
		\vspace{-0.3cm}
		\setlength{\abovecaptionskip}{-0.04cm}
		\subfigure[Clean images]{
			\label{Fig. 6(a)}
			\includegraphics[width=0.19\textwidth,clip,trim=0 0 0 0]{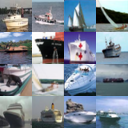}}
		\subfigure[Poisoned images of BadNets]{
			\label{Fig. 6(b)}
			\includegraphics[width=0.19\textwidth,clip,trim=0 0 0 0]{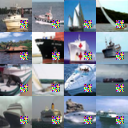}}
		\subfigure[Poisoned images of CLB]{
			\label{Fig. 6(c)}
			\includegraphics[width=0.19\textwidth,clip,trim=0 0 0 0]{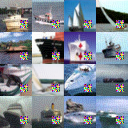}}
		\subfigure[Poisoned images of DeNeB]{
			\label{Fig. 6(d)}
			\includegraphics[width=0.19\textwidth,clip,trim=0 0 0 0]{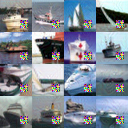}}
		\subfigure[Our poisoned images]{
			\label{Fig. 6(e)}
			\includegraphics[width=0.19\textwidth,clip,trim=0 0 0 0]{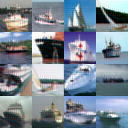}}
		\caption{Clean images and poisoned images from CIFAR10. These images from SVHN are posted on the supplementary.}
		\vspace{-0.4cm}
		\label{Fig. 6}
	\end{figure*}
	
	%
	%
	%
	%
	%
	%
	%
	
	\section{Experiment evaluation}
	
	\subsection{Experiment setup}
	\subsubsection{Victim network and dataset:}
	We implement our poisoning on CIFAR10 \cite{krizhevsky2009learning} and SVHN \cite{netzer2011reading}, which are widely used in semi-supervised learning. CIFAR10 contains 50,000 training images and 10,000 testing images from 10 classes. SVHN consists of 73257 training images and 26032 testing images of house digits from 10 classes. Moreover, CIFAR10 is trained on CNN13 \cite{tarvainen2017mean} and SVHN is trained on WideResNet-28-2 \cite{zagoruyko2016wide}. To implement semi-supervised learning, in CIFAR10, only 4000 training images are labeled, i.e., $N=4000$. In SVHN, only 1000 images are labeled, i.e., $N=1000$.
	
	\subsubsection{SSL algorithms:} We select some representative SSL algorithms from three types introduced in Section 2.3: consistency regularization is PI-Model \cite{rasmus2015semi}, MeanTeacher \cite{tarvainen2017mean}, VAT \cite{miyato2018virtual}, and ICT \cite{verma2019interpolation}, pseudo-labeling is PseudoLabel \cite{lee2013pseudo}, pseudo-labeling with consistency regularization is FixMatch \cite{sohn2020fixmatch}. Some of these algorithms are old and do not perform well, while others are recently proposed and have outstanding performance, which can fully verify the generality of our poisoning. The implementations of these SSL algorithms on SVHN and CIFAR10 come from the public Pytorch open source codes, \cite{suzuki2020consistency} and \cite{SSLset}, respectively.
	
	\subsubsection{Baseline poisonings:} We compare our poisoning with BadNets-C \cite{gu2017badnets}, CLB \cite{turner2019label}, and DeNeB-C \cite{yan2021deep} , which have been described in detail in Section 3. 
	Their backdoor patterns are all the same $8\times8$ pixel block located at the position (20, 20) of the image, as shown in Fig. \ref{Fig. 6(b)} \ref{Fig. 6(c)}, and \ref{Fig. 6(d)}. The target class is all 8.

	\subsubsection{Poisoning setup:} In our experiments, except that the surrogate dataset for victim dataset CIFAR10 is CIFAR10, and the surrogate dataset for victim dataset SVHN is SVHN, the other experimental settings are the same. The surrogate network and the backdoor pattern generator are WideResNet-28-2 \cite{zagoruyko2016wide} and UNet \cite{ronneberger2015u}, respectively. 
	The target class is 8, the number of poisons is 500, $\lambda_{ins}$ is 0.05, $\epsilon$ is 27 which indicates pixel perturbation maximum.
	\subsubsection{Evaluation metrics:} Evaluation metrics include three: the accuracy of clean examples (CA), the attack success rate (ASR) of the backdoor, and the imperceptibility (IMP) of backdoor patterns.
	
	\textbf{CA}: Backdoor poisoning should not degrade the accuracy of the SSL model, that is, the CA on the poisoned model should be close to the CA on the unpoisoned model.
	
	\textbf{ASR}: In the inference stage, the probability that the examples with backdoor patterns added are misclassified by the poisoned model into the target class, higher ASR means better poisoning performance.
	
	\textbf{IMP}: The more imperceptible backdoor patterns are, the better they can evade the detection of the victim. We quantify imperceptibility by computing the distance between clean images and poisoned images by PSNR \cite{hore2010image}, SSIM \cite{hore2010image}, and L-$\infty$ norm. The larger the PSNR, the closer the SSIM is to 1, and the smaller the L-$\infty$ norm, the better the imperceptibility.

	\subsection{Poisoning performance}
	Let's first verify that the poisoned unlabeled examples and gradient matching work in the trained-from-scratch SSL model. In the SSL model of the $k$th epoch, the loss of poisoned examples taking the target label as the label is:
	\begin{equation}
		\mathcal{D}=\ell({y^t},\mathcal{F}({\mathcal{P}_{tr}}({\mu _b}\mathcal{U});{\theta _k}))
	\end{equation}
	To more accurately reflect that predictions of backdoor examples are far away from the clean classes and close to the target class, we adopt relative distance to define the degree $\mathcal{C}$ of gradient matching.
	\begin{equation}
		\mathcal{C} = {\mathbb{E}_{(x,y) \in \mathcal{X}_{val}}}\frac{{{{\left\| {{\nabla _{{\theta _k}}}\ell({y^t},\mathcal{F}({\mathcal{P}_{val}}(x);{\theta _k})) - {\nabla _{{\theta _k}}}\ell({y^t},\mathcal{F}({x^t};{\theta _k}))} \right\|}^2}}}{{{{\left\| {{\nabla _{{\theta _k}}}\ell(y,\mathcal{F}({\mathcal{P}_{val}}(x);{\theta _k})) - {\nabla _{{\theta _k}}}\ell({y^t},\mathcal{F}({x^t};{\theta _k}))} \right\|}^2}}}
	\end{equation}
	where the upper term calculates the gradient distance between the backdoor example $\mathcal{P}_{val}(x)$ taking the target label as the label and the target example $x^t$ taking the target label as the label. The lower term calculates the gradient distance between the backdoor example taking the correct label as the label and the target example taking the target label as the label.
	
\begin{figure}[ht]
		\vspace{-0.2cm}
		\setlength{\abovecaptionskip}{-0.05cm}
		\subfigure[PseudoLabel]{
			\label{Fig. 7(a)}
			\includegraphics[width=0.15\textwidth,clip,trim=0 10 40 60]{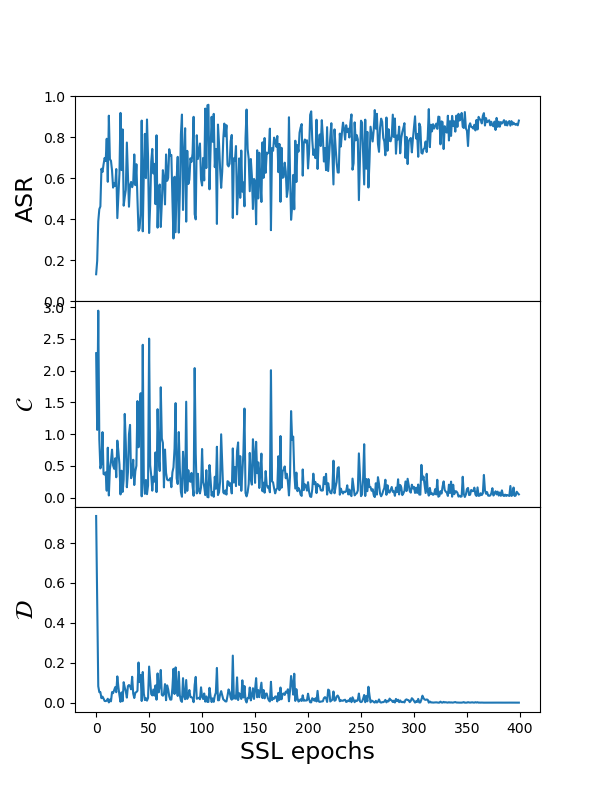}}
		\subfigure[PI-Model]{
			\label{Fig. 7(b)}
			\includegraphics[width=0.15\textwidth,clip,trim=0 10 40 60]{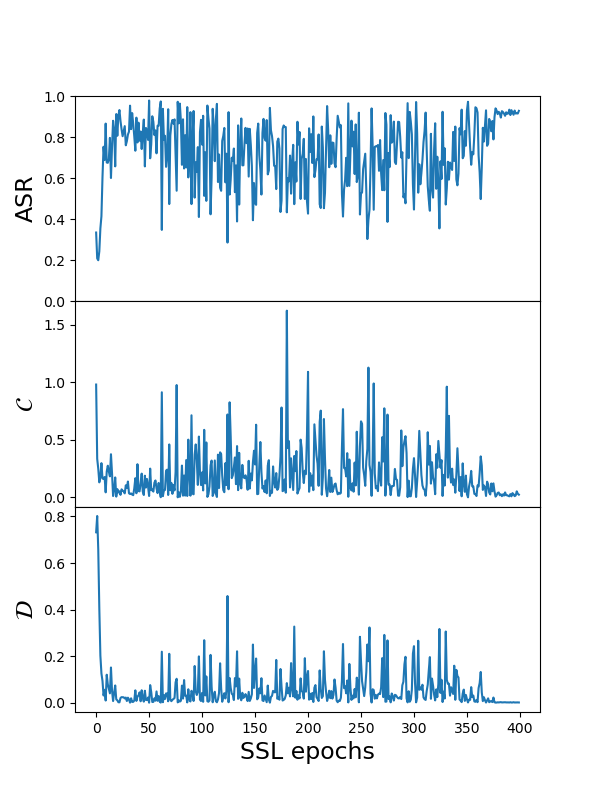}}
		\subfigure[MeanTeacher]{
			\label{Fig. 7(c)}
			\includegraphics[width=0.15\textwidth,clip,trim=0 10 40 60]{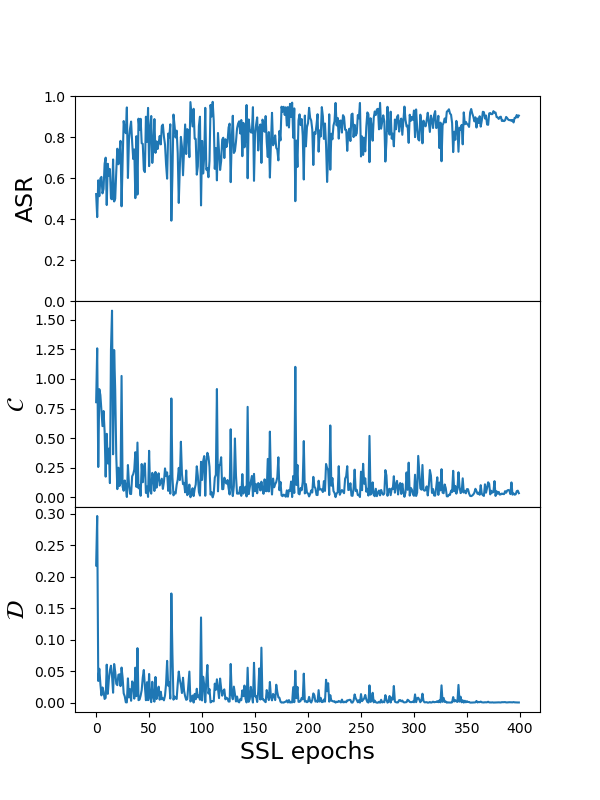}}
		\subfigure[VAT]{
			\label{Fig. 7(d)}
			\includegraphics[width=0.15\textwidth,clip,trim=0 10 40 60]{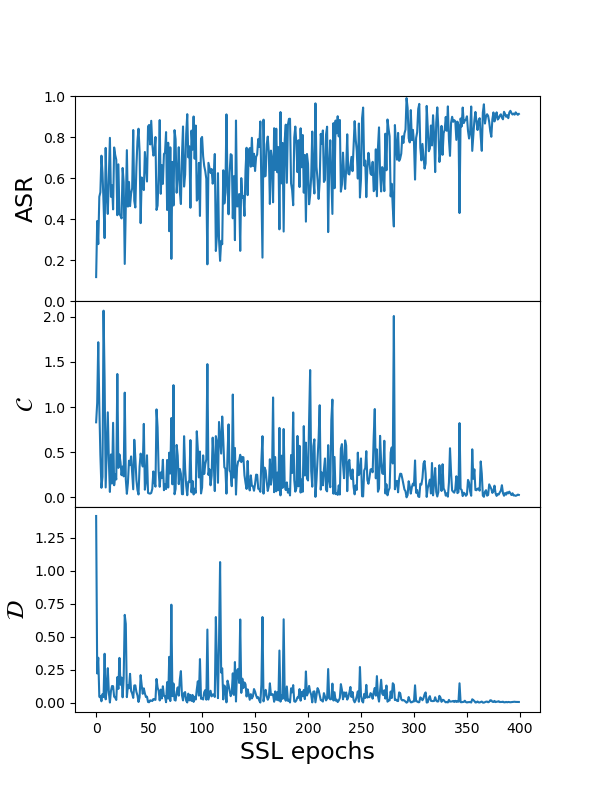}}
		\subfigure[ICT]{
			\label{Fig. 7(e)}
			\includegraphics[width=0.15\textwidth,clip,trim=0 10 40 60]{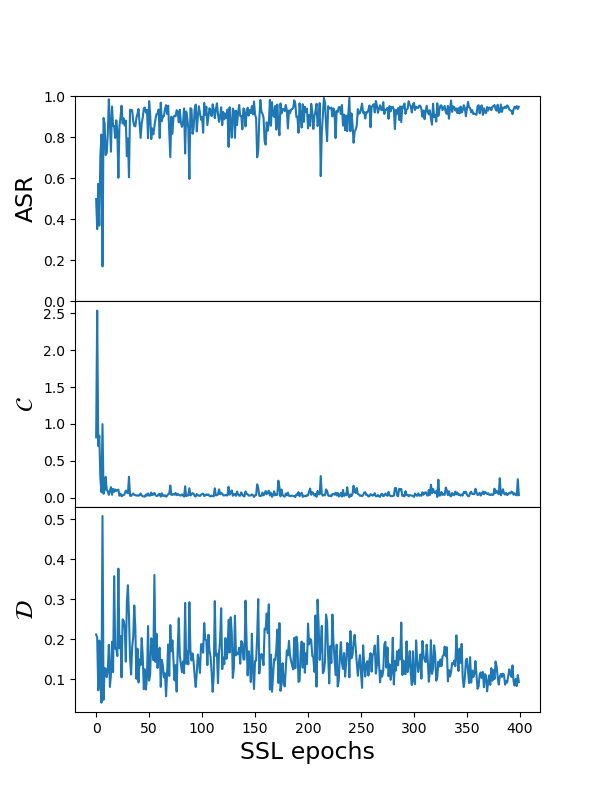}}
		\subfigure[FixMatch]{
			\label{Fig. 7(f)}
			\includegraphics[width=0.15\textwidth,clip,trim=0 10 40 60]{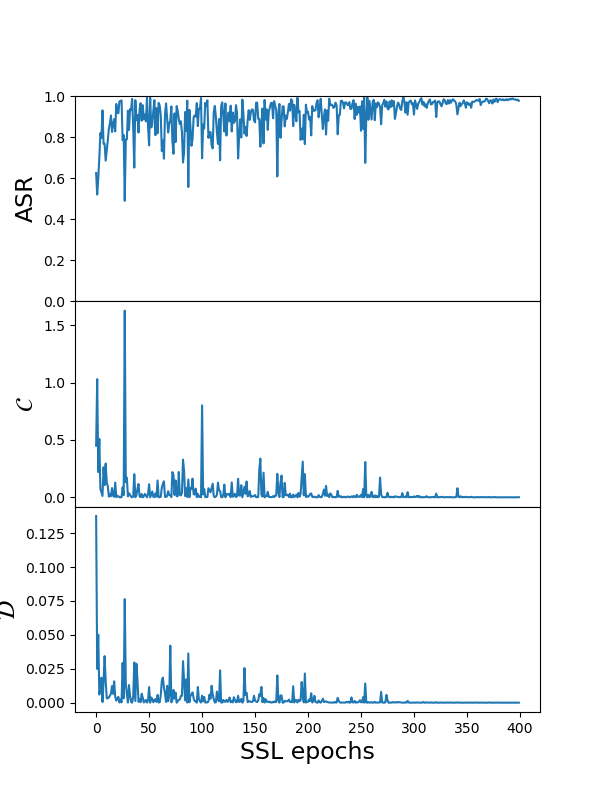}}
		\caption{The evolutions of $\mathcal{C}$, $\mathcal{D}$, and ASR with increasing epochs on trained-from-scratch SSL model CNN13.}
		\vspace{-0.5cm}
		\label{Fig. 7}
	\end{figure}

	Since poisoned unlabeled examples are all from the target class, the SSL model will correctly classify them as the target class $y^t$, so $\mathcal{D}$ will gradually decrease, as shown in the bottom of Fig. \ref{Fig. 7}. As a result, the gradient distance $\mathcal{C}$ between backdoor examples and target examples also gradually decreases, as shown in the middle of Fig. \ref{Fig. 7}. This brings about a gradual increase in ASR, as shown in the top of Fig. \ref{Fig. 7}. In Table \ref{Table 1}, we present the final poisoning results, from which we can draw three conclusions.
	


	%
	%
	%
	First, SSL CAs are significantly higher than SL CAs, which means that the learning of SSL algorithms on unlabeled examples improves the model accuracy. Moreover, when poisonings are implemented, the CAs of poisoned models do not show significant degradations compared to CAs of clean models, which means that unlabeled backdoor poisoning can be achieved without degenerating model accuracy.
	
	Second, on these SSL algorithms, the ASRs of our poisoning are much higher than those of several baseline poisonings. As listed in Table \ref{Table 1}, on CIFAR10, although these poisonings achieve certain ASRs, the highest is only $68.56\%$, while our lowest is $87.13\%$. On SVHN, baseline poisonings fail on all SSL algorithms except FixMatch which has the ASR of $75.47\%$. In contrast, our poisoning can be applied to these SSL algorithms. Although ASRs of our poisoning are lower on SVHN than on CIFAR10, the imperceptibility of backdoor patterns is better. If the adversary is willing to sacrifice imperceptibility, it will bring an increase in ASRs. In addition, different SSL algorithms have different vulnerabilities to our poisoning. On FixMatch, our poisoning performs the best, while on VAT, the ASR is the lowest. This may be because VAT considers adversarial perturbations as image augmentations and implements adversarial training to ensure that unlabeled examples are resistant to these adversarial perturbations. The backdoor patterns we craft are similar to adversarial perturbations, so adversarial training improves the model’s ability to resist poisoned examples.
	
	Finally, thanks to the imperceptibility design of our poisoning, poisoned images look very similar to clean images, as shown in Fig. \ref{Fig. 6}. In contrast, the poisoned images of baseline poisonings have obvious backdoor squares, which are easily detected by victims as suspicious. Quantitatively, as listed in Table \ref{Table 1}, our PNSR, SSIM and L-$\infty$ norm all significantly outperform those of these schemes. To sum up, our poisoning well achieves the poisoning target in Eq. \ref{Eq. 2}.

	\subsection{Ablation study}
	To focus on the impact of varying hyperparameters or situations on poisoning performance, the evaluations in this section are performed on CIFAR10 trained with FixMatch. 
	Poisoning on other target classes are posted on the supplementary.
	\begin{table}
		\vspace{-0.cm}
		\setlength{\abovecaptionskip}{-0.01cm}
		\centering
		\caption{Evaluation across network architectures. SimNet is a simple network we build that only consists of one convolutional layer ($3\times64$) for down-sampling and one deconvolutional layer ($64\times3$) for up-sampling.}
		\scalebox{0.89}{\begin{tabular}{c|c|c|c|cll} 
				\hline
				\multirow{2}{*}{Generator} & \multirow{2}{*}{Surrogate Nets} & \multirow{2}{*}{CA} & \multirow{2}{*}{ASR} & \multicolumn{3}{c}{IMP}  \\
				&                                 &                     &                      & PSNR  & SSIM   & L-$\infty$    \\ 
				\hline
				\multirow{3}{*}{SimNet}    & LeNet                           & 93.72               & 99.33                & 21.72 & 0.6950 & 27.00   \\
				& CNN13                           & 93.73               & 90.40                & 26.54 & 0.8703 & 26.89   \\
				& WideResNet-28-2                 & 93.65               & 94.39                & 26.63 & 0.8966 & 26.87   \\ 
				\hline
				\multirow{3}{*}{UNet}      & LeNet                           & 93.77               & 94.37                & 24.65 & 0.8057 & 27.00   \\
				& CNN13                           & 93.86               & 95.24                & 31.43 & 0.9423 & 24.88   \\
				& WideResNet-28-2                 & 93.83               & 97.12                & 31.34 & 0.9515 & 24.51   \\
				\hline
		\end{tabular}}
		\label{Table 2}
		\vspace{-0.2cm}
	\end{table}
	\subsubsection{Evaluation across network architectures}
	We evaluate the impact of different architectures of generators and surrogate networks on the ASR and imperceptibility.
	Alternative generators include SimNet and UNet. Alternative surrogate networks are LeNet \cite{lecun1998gradient}, CNN13, and WideResNet-28-2. The poisoning results are listed in Table \ref{Table 2}. Comparing SimNet and UNet, it can be seen that because the network is too simple, it is more difficult for SimNet to explore imperceptible backdoor patterns. Although a higher ASR is obtained on LeNet, imperceptibility is greatly sacrificed. On CNN13 and WideResNet-28-2, UNet achieves higher ASR and better imperceptibility. Comparing these surrogate networks, WideResNet-28-2 is larger in scale, better at exploring the least perceptible backdoor patterns, and obtaining higher ASR.
	\subsubsection{Evaluation across perturbation budgets $\epsilon$}
	Although a higher perturbation budget $\epsilon$ leads to more significant changes in individual pixels, it allows us more space to search for imperceptible backdoor patterns and leads to higher ASRs. As shown in Table \ref{Table 4}, at $\epsilon=7$, although the maximum value of pixel perturbation is only 7, the imperceptibility of backdoor patterns does not bring more significant improvement than at $\epsilon=27$, while the ASR significantly drops, from $97.12\%$ to $64.40\%$. On the other hand, when $\epsilon=54$, the maximum value of pixel perturbation is improved to 32.32 compared to at $\epsilon=27$, while the ASR is only improved by $1.51\%$. Considering ASR and imperceptibility, $\epsilon=27$ is more suitable.
	\subsubsection{Evaluation across other situations}
	\begin{figure}[ht]
		\vspace{-0.6cm}
		\setlength{\abovecaptionskip}{-0.05cm}
		\subfigure[Poisoned images of CLB]{
			\label{Fig. 8(a)}
			\includegraphics[width=0.19\textwidth,clip,trim=0 0 0 0]{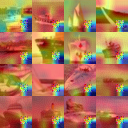}}
		\subfigure[Our poisoned images]{
			\label{Fig. 8(b)}
			\includegraphics[width=0.19\textwidth,clip,trim=0 0 0 0]{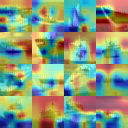}}
		\caption{Grad-CAM \cite{selvaraju2017grad} visualiztion.}
		\vspace{-0.6cm}
		\label{Fig. 8}
	\end{figure}
	In this section, we consider three situations of our poisoning. 
	
	S1: Assume that the adversary does not know that the victim dataset is CIFAR10, but only knows the target class and dataset distribution. He then employs CIFAR100, which has a similar distribution to CIFAR10, as a surrogate dataset, and replace examples of a certain class with examples from the target class. 
	
	S2: Assume that the adversary does not know the labeling situation of the examples in the victim training set. The poisoned unlabeled examples are then labeled proportionally.
	
	S3: Assume that the poisoned unlabeled examples come from different classes.
	
	The poisoning results are listed in Table \ref{Table 5}. In the S1 situation, since the distributions of CIFAR100 and CIFAR10 are similar, backdoor patterns crafted on CIFAR100 can be migrated to CIFAR10, obtaining an ASR of $94.25\%$. In the S2 situation, since we only poison examples from the target class, even if these examples are correctly labeled, it will not impact the ASR. The poisoning failure (the ASR is only $8.79\%$) in the S3 situation further validates the finding in Section 3: label-inconsistent backdoor poisoning is much more difficult to use for unlabeled examples.
	\begin{table}
		\centering
		\vspace{-0.2cm}
		\setlength{\abovecaptionskip}{-0.01cm}
		\caption{Evaluation across perturbation budgets $\epsilon$}
		\begin{tabular}{c|c|c|c|ccc} 
			\hline
			\multirow{2}{*}{$\epsilon$} & \multirow{2}{*}{SSL CA} & \multirow{2}{*}{CA} & \multirow{2}{*}{ASR} & \multicolumn{3}{c}{IMP}  \\
			&                         &                     &                      & PSNR  & SSIM   & L-$\infty$   \\ 
			\hline
			7                  & \multirow{3}{*}{93.56}  & 93.62               & 64.40                & 33.12 & 0.9502 & 7.00    \\
			27                 &                         & 93.83               & 97.12                & 31.34 & 0.9515 & 24.51   \\
			54                 &                         & 93.78               & 98.63                & 31.57 & 0.9497 & 32.32   \\
			\hline
		\end{tabular}
		\label{Table 4}
		\vspace{-0.3cm}
	\end{table}
	\begin{table}
		\setlength{\abovecaptionskip}{-0.01cm}
		\centering
		\caption{Evaluation across other situations}
		\begin{tabular}{c|c|c|c|ccc} 
			\hline
			\multirow{2}{*}{Situation} & \multirow{2}{*}{SSL CA} & \multirow{2}{*}{CA} & \multirow{2}{*}{ASR} & \multicolumn{3}{c}{IMP}      \\
			&                         &                     &                      & PSNR  & SSIM   & L-$\infty$  \\ 
			\hline
			S1                         & \multirow{3}{*}{93.56}  & 93.69               & 94.25                & 30.25 & 0.9389 & 24.59       \\
			S2                         &                         & 93.75               & 96.89                & 31.34 & 0.9515 & 24.51       \\
			S3                         &                         & 93.86               & 8.79                 & 31.34 & 0.9515 & 24.51       \\
			\hline
		\end{tabular}
		\label{Table 5}
		\vspace{-0.3cm}
	\end{table}
	\subsection{Defense evaluation}
	The section evaluates that our poisoning can bypass five representative defenses from four types mentioned in Section 2.2, including Activation Cluster \cite{chen2018detecting}, Neural Cleanse \cite{wang2019neural}, Fine-pruning \cite{liu2018fine}, STRIP \cite{gao2019strip}, and DePuD \cite{yan2021deep}. The DePuD is posted below, and the other four are posted on the supplementary.
	\subsubsection{DePuD}
\begin{figure}[ht]
		\vspace{-0.3cm}
		\setlength{\abovecaptionskip}{-0.05cm}
		\subfigure[]{
			\label{Fig. 9(a)}
			\includegraphics[width=0.22\textwidth,clip,trim=10 0 20 20]{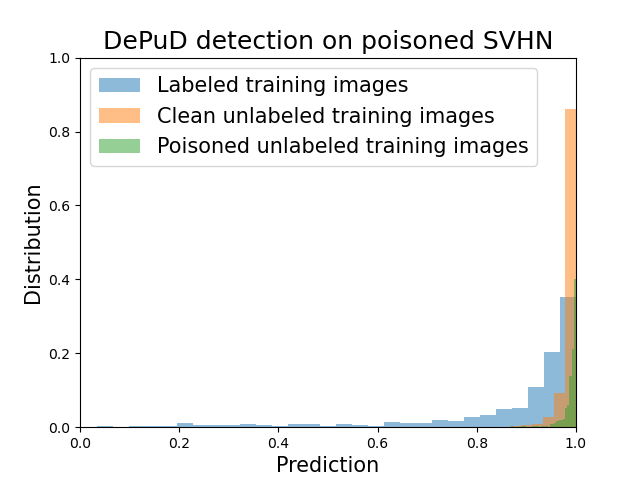}}
		\subfigure[]{
			\label{Fig. 9(b)}
			\includegraphics[width=0.22\textwidth,clip,trim=10 0 20 20]{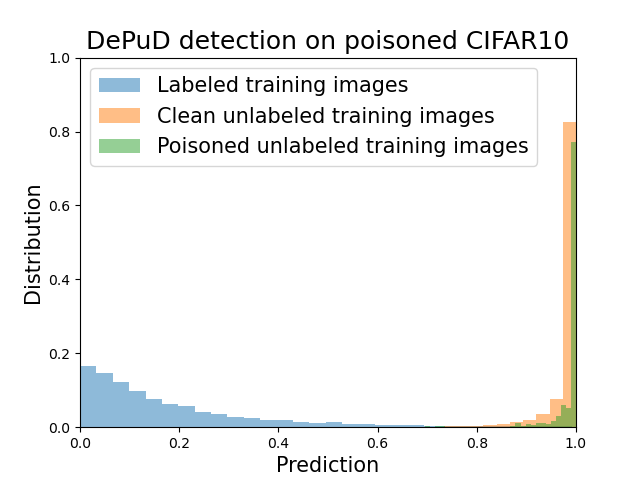}}
		\caption{DePuD detection.}
		\label{Fig. 9}
		\vspace{-0.3cm}
	\end{figure}

	DePuD is proposed to detect poisoned unlabeled examples. First, all the training examples are divided into two categories according to whether they are labeled. The labeled ones are assigned the label 0, and the unlabeled ones are assigned the label 1. These examples are then classified using a heavy regularization model. If poisoned examples have significant backdoor patterns, the predictions will be extremely close to 1, and the separation from clean unlabeled examples will appear, so that it is detected as abnormal. DePuD works for CLB, as shown in Fig. \ref{Fig. 8(a)}, backdoor patterns can be detected prominently in the lower right corner. However, the backdoor patterns of our poisoning are imperceptible, it is difficult to be captured by the heavy regularization model, as shown in Fig. \ref{Fig. 8(b)}. Moreover, as shown in Fig. \ref{Fig. 9}, clean unlabeled examples almost overlap with poisoned unlabeled examples, which means that DePuD cannot separate the poisoned unlabeled examples. 
	%
	%
	%
	
	\section{Conclusion}
	This paper is the first to investigate the vulnerability of unlabeled examples of trained-from-scratch SSL models to backdoor poisoning, revealing the flaws in the security design of SSL algorithms. We first find that label-inconsistent backdoor poisoning 
	cannot be used for unlabeled examples due to the opposition to the SSL algorithms that strive to correctly learn unlabeled examples. Thus, for unlabeled examples, poisoning only is implemented on examples from the target class. Based on this, we propose a zero-knowledge and imperceptible backdoor poisoning. Experiments show that our poisoning achieves state-of-the-art attack success rates when bypassing various defenses.
	\bibliography{acmart}
	\bibliographystyle{aaai23} 
	
	\section*{Appendix}
	\appendix
	\section{SVHN images}
	In Fig. \ref{Fig. A4}, we show clean images and poisoned images on SVHN. It can be seen that our poisoned images are visually very similar to the clean images, and the victim is difficult to detect.
	\section{Evaluation across target classes}
	We select other classes to act as target classes. The poisoning results are listed in Table \ref{Table A3}. First, likewise, poisoning on other target classes does not degrade the model accuracy. Secondly, it can be seen that the poisoning difficulty of different target classes is different. For example, on the target class Bird, the ASR can reach $99.52\%$, while on Truck, the ASR is lower, $81.88\%$. Of course, we also can sacrifice a little backdoor pattern imperceptibility to improve the ASR.

	\section{Defense evaluation }
	We evaluate our poisoning on Activation Cluster, Neural Cleanse, Fine-pruning, and STRIP.
	\subsection{Activation Cluster}
	The process of Activation Cluster is to input all training examples into the already trained victim model, thereby obtaining the activation of these examples in the last hidden layer. These activations are then divided into different clusters based on their labels. Finally, it is determined whether there are poisoned examples by detecting the abnormality of these clusters. However, our backdoor poisoning does not rely on labels and poisons only unlabeled examples. Thus, poisoned unlabeled examples cannot be divided into different clusters based on labels. Thus, our poisoning can naturally bypass Activation Cluster detection.
	\subsection{Neural Cleanse} 
\begin{figure}[ht]
		\vspace{-0.4cm}
		\setlength{\abovecaptionskip}{-0.05cm}
		\subfigure[]{
			\label{Fig. A1(a)}
			\includegraphics[width=0.214\textwidth,clip,trim=30 0 40 20]{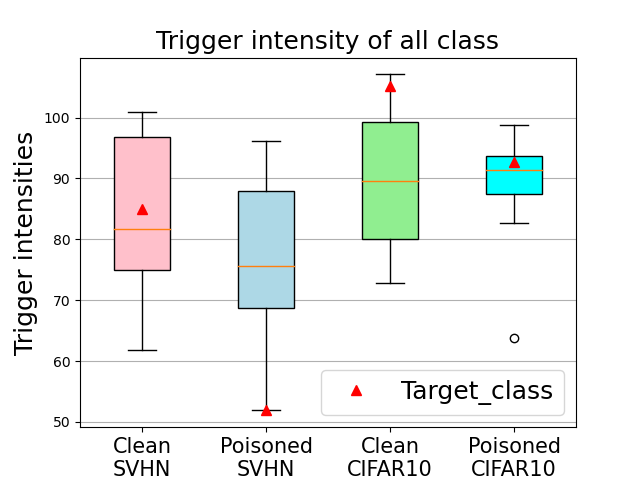}}
		\subfigure[]{
			\label{Fig. A1(b)}
			\includegraphics[width=0.24\textwidth,clip,trim=0 0 0 0]{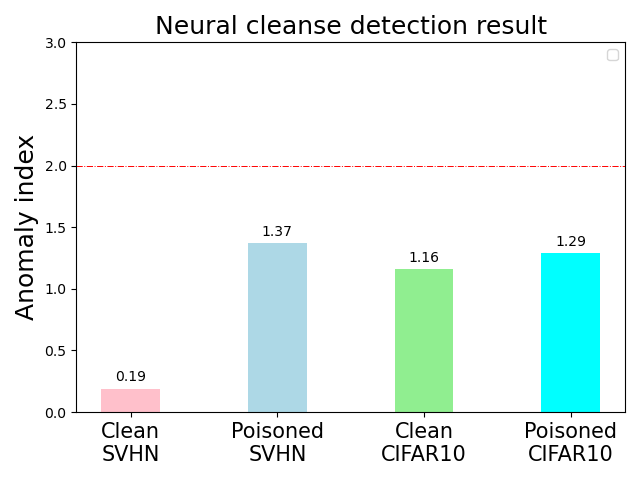}}
		\label{Fig. A1}
		\caption{Neural Cleanse detection.}
		\vspace{-0.2cm}
	\end{figure}

On potentially poisoned models, reverse-engineer the minimum-intensity backdoor triggers for all classes. They then determine whether a certain class is the target class based on the prior knowledge that the target class injected into the backdoor has a trigger with abnormally small intensity, i.e., $anomaly \ index > 2$. In semi-supervised learning, the defender may not have enough labeled examples for more accurate reverse engineering, but we assume the most stringent condition that the defender has enough labeled examples. However, even so, as shown in the Fig. \ref{Fig. A1(a)}, the trigger intensity of the target class injected into the backdoor is not significant outliers, and $anomaly \ index$ is all less than 2 (Fig. \ref{Fig. A1(b)}). Thus, our poisoning can successfully bypass Neural Cleanse detection. We think this is because reverse engineering in Nerucal Cleanse detection relies on classification layers, whereas our poisoning is gradient matching that controls the entire network.
	
	\subsection{Fine-pruning}
\begin{figure}[ht]
		\vspace{-0.4cm}
		\setlength{\abovecaptionskip}{-0.05cm}
		\subfigure[]{
			\label{Fig. A2(a)}
			\includegraphics[width=0.22\textwidth,clip,trim=10 0 0 20]{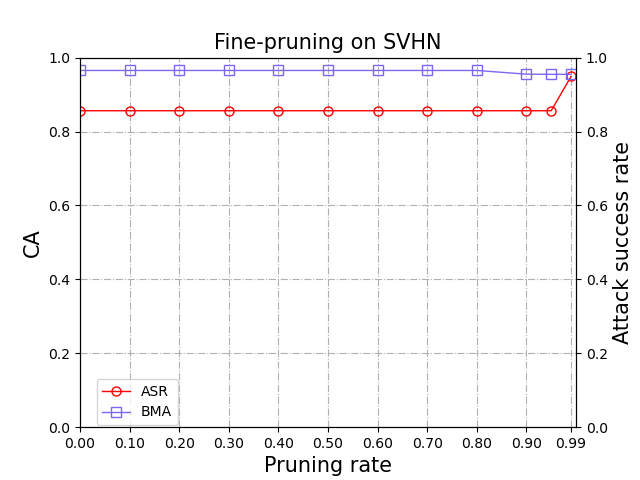}}
		\subfigure[]{
			\label{Fig. A2(b)}
			\includegraphics[width=0.22\textwidth,clip,trim=10 0 0 20]{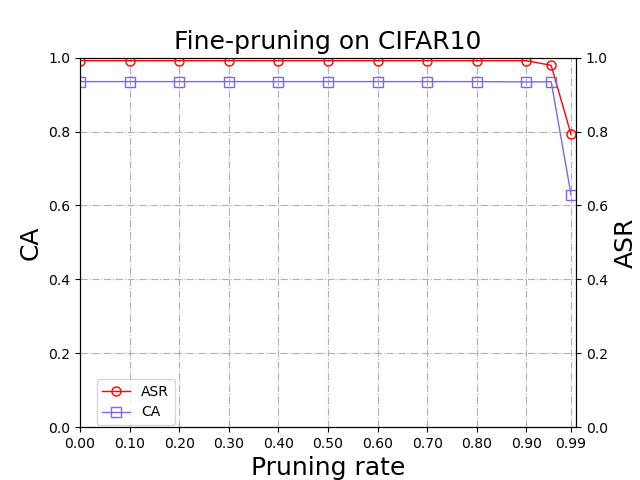}}
		\caption{Fine-pruning.}
		\label{Fig. A2}
		\vspace{-0.5cm}
	\end{figure}

\begin{figure*}[ht]
		\vspace{-0.3cm}
		\setlength{\abovecaptionskip}{-0.04cm}
		\subfigure[Clean images]{
			\label{Fig. A4(a)}
			\includegraphics[width=0.19\textwidth,clip,trim=0 0 0 0]{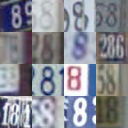}}
		\subfigure[Poisoned images of BadNets]{
			\label{Fig. A4(b)}
			\includegraphics[width=0.19\textwidth,clip,trim=0 0 0 0]{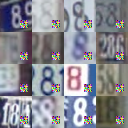}}
		\subfigure[Poisoned images of CLB]{
			\label{Fig. A4(c)}
			\includegraphics[width=0.19\textwidth,clip,trim=0 0 0 0]{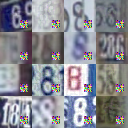}}
		\subfigure[Poisoned images of DeNeB]{
			\label{Fig. A4(d)}
			\includegraphics[width=0.19\textwidth,clip,trim=0 0 0 0]{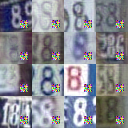}}
		\subfigure[Our poisoned images]{
			\label{Fig. A4(e)}
			\includegraphics[width=0.19\textwidth,clip,trim=0 0 0 0]{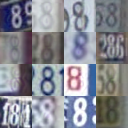}}
		\caption{Clean images and poisoned images on SVHN.}
		\vspace{-0.4cm}
		\label{Fig. A4}
	\end{figure*}

	This is a blind defense strategy, instead of detecting whether the model or example is poisoned, it uses pruning and fine-tuning for any model to try to eliminate possible backdoors. Specifically, clean examples are first fed into the model, and then $\alpha$\% (i.e., Pruning rate) of neurons with minimal activation are dormant by pruning, thereby attempting to remove possible backdoors. Fine-tuning is then used to compensate for the degradation of clean example accuracy caused by pruning. As shown in Fig. \ref{Fig. A2}, even with a pruning rate of 90\%, there is no significant drop in the ASR. When the pruning rate is 99\%, on CIFAR10, the CA drops to 80\%, and the ASR is still 62\%. An interesting phenomenon is that on SVHN, the ASR increases significantly, which may be because the excessive pruning makes the model's ability to distinguish clean examples weakened, which makes it easier to be misclassified as the target class once backdoor patterns are added. 
	\begin{table}
		\vspace{-0.02cm}
		\centering
		\caption{Evaluation across target classes}
		\begin{tabular}{c|c|c|c|cll} 
			\hline
			\multirow{2}{*}{Target class}   & \multirow{2}{*}{SSL CA} & \multirow{2}{*}{CA}        & \multirow{2}{*}{ASR}       & \multicolumn{3}{c}{IMP}                     \\
			&                         &                            &                            & \multicolumn{1}{l}{PSNR}  & SSIM   & L-$\infty$  \\ 
			\hline
			Airplane                        & \multirow{10}{*}{93.56} & 93.49                      & 99.15                      & 33.29                     & 0.9518 & 24.26  \\
			\multicolumn{1}{l|}{Automobile} &                         & \multicolumn{1}{l|}{93.44} & \multicolumn{1}{l|}{89.22} & 30.21 & 0.9490 & 24.88  \\
			Bird                            &                         & 93.76                      & 99.52                      & 33.08                     & 0.9492 & 23.99  \\
			Cat                             &                         & 93.95                      & 99.95                      & 33.33                     & 0.9655 & 20.06  \\
			Deer                            &                         & 93.78                      & 98.18                      & 32.51                     & 0.9560 & 22.18  \\
			Dog                             &                         & 93.81                      & 91.08                      & 32.32                     & 0.9634 & 23.81  \\
			Frog                            &                         & 93.89                      & 95.64                      & 31.47                     & 0.9519 & 23.22  \\
			Horse                           &                         & 93.84                      & 94.11                      & 32.36                     & 0.9653 & 22.91  \\
			Ship                            &                         & 93.83                      & 97.12                      & 31.34                     & 0.9515 & 24.51  \\
			Truck                           &                         & 93.93                      & 81.88                      & 32.65                     & 0.9765 & 23.29  \\
			\hline
		\end{tabular}
		\label{Table A3}
		\vspace{-0.2cm}
	\end{table}
	\subsection{STRIP}
\begin{figure}[ht]
		\vspace{-0.2cm}
		\setlength{\abovecaptionskip}{-0.05cm}
		\subfigure[]{
			\label{Fig. A3(a)}
			\includegraphics[width=0.23\textwidth,clip,trim=40 0 40 20]{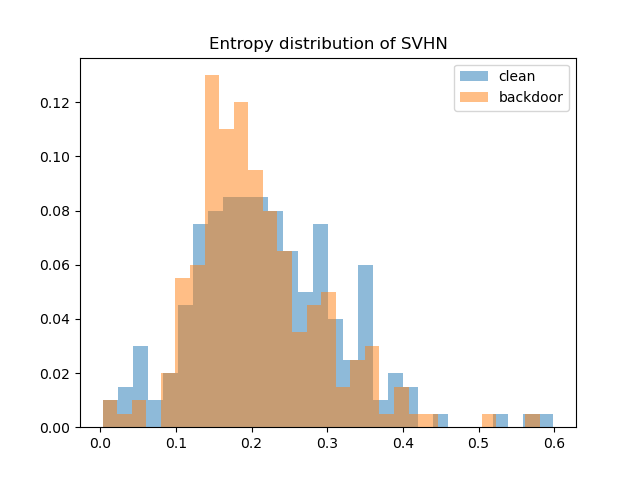}}
		\subfigure[]{
			\label{Fig. A3(b)}
			\includegraphics[width=0.23\textwidth,clip,trim=40 0 40 20]{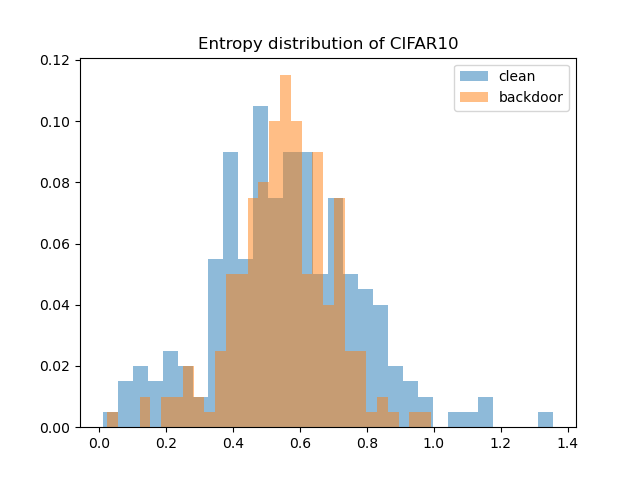}}
		\caption{STRIP detection.}
		\vspace{-0.5cm}
		\label{Fig. A3}
	\end{figure}

	STRIP is deployed in the model inference stage. Before the testing example is fed to the model, it is synthesized with a set of pre-prepared clean examples. Then these obtained synthesized examples are fed into the model for prediction. If the entropy of their prediction results is abnormally small, it is determined that the testing example is a backdoor example, and the model is poisoned. As shown in Fig. \ref{Fig. A3}, we show the entropy distribution for 500 testing examples and 500 backdoor examples, and it can be seen that the distributions almost coincide. The entropy of the backdoor examples does not exhibit abnormally small property. Thus, our backdoor poisoning can bypass STRIP detection. 
\end{document}